\documentclass[aps,prl,showpacs,preprintnumbers,superscriptaddress,nofootinbib,groupedaddress]{revtex4} 

\usepackage{color}
\usepackage{graphicx}
\usepackage{amsmath}
\usepackage{amssymb}
\usepackage[colorlinks=true,citecolor=darkred,urlcolor=darkred, pdfborder={0 0 0}]{hyperref}
\usepackage[normalem]{ulem}
\definecolor{darkred}{rgb}{0.6,0,0}

\definecolor{linkcolor}{rgb}{0,0,0.5}

\def\gsim{\raise0.3ex\hbox{$\;>$\kern-0.75em\raise-1.1ex\hbox{$\sim\;$}}}
\def\lsim{\raise0.3ex\hbox{$\;<$\kern-0.75em\raise-1.1ex\hbox{$\sim\;$}}}

\def\beqn#1{\begin{equation}\label{#1}}
\def\eeqn{\end{equation}}

\def\beqa#1{\begin{eqnarray}\label{#1}}
\def\eeqa{\end{eqnarray}}

\def\Z2{$\mathcal{Z_2}$}

\def\vev#1{\left\langle #1\right\rangle}
 
\newcommand {\ignore}[1]{}

\newcommand{\sm}{{Standard Model }}
\newcommand{\AddrAHEP}{%
  AHEP Group, Institut de F\'{i}sica Corpuscular --
  CSIC/Universitat de Val\`{e}ncia, Parc Cient\'ific de Paterna.\\
 C/ Catedr\'atico Jos\'e Beltr\'an, 2 E-46980 Paterna (Valencia) - SPAIN}
 \def\one{\ensuremath{\mathbf{1}}}
 \def\two{\ensuremath{\mathbf{2}}}
 \def\three{\ensuremath{\mathbf{3}}}
 	
 \def\five{\ensuremath{\mathbf{5}} }

\def\SM{$\mathrm{SU(3)_c \otimes SU(2)_L \otimes U(1)_Y}$ }

\begin{document}
\bibliographystyle{unsrt}
\title{SO(3) family symmetry and  axions}
\author{Mario Reig}
\email{mario.reig@ific.uv.es}
\affiliation{\AddrAHEP}

\author{Jos\'e W.F. Valle}
\email{valle@ific.uv.es}
\affiliation{\AddrAHEP}

\author{Frank Wilczek} \email{wilczek@mit.edu} \affiliation{Center for
  Theoretical Physics, Massachusetts Institute of Technology, Cambridge, Massachusetts 02139 USA}
\affiliation{Tsung-Dao Lee Institute and Wilczek Quantum Center,  Shanghai 200240, China}
\affiliation{Department of Physics, Stockholm University,  Stockholm SE-106 91 Sweden}
\affiliation{Department of Physics and Origins Project, Arizona State University, Tempe, Arizona 25287, USA}
\date{\today}


\newcommand {\black} {\color{black}}
\newcommand {\blue} {\color{blue}}
\newcommand {\cyan} {\color{cyan}}
\newcommand {\green} {\color{green}}
\newcommand {\yellow} {\color{yellow}}
\newcommand {\magenta} {\color{magenta}}
\newcommand {\red} {\color{red}}


\begin{abstract}
  Motivated by the idea of comprehensive unification, we study a
  gauged SO(3) flavor extension of the extended \sm, including right-handed neutrinos and a Peccei-Quinn
  symmetry with simple charge assignments.  The model accommodates the
  observed fermion masses and mixings and yields a characteristic,
  successful relation among them. The Peccei-Quinn symmetry is an essential
  ingredient.  \end{abstract}

\preprint{MIT-CTP/5003}

\maketitle

\section{Introduction and motivation}
\label{motivation}
For all its success, the \sm has many loose ends and shortcomings.  It leaves unexplained the threefold family replication, the observed pattern of quark and lepton masses and mixings, and the lack of CP violation in the strong interaction~\cite{Peccei:1977hh,Weinberg:1977ma,Wilczek:1977pj}.  It
does not account for the cosmological dark matter~\cite{Bertone:2004pz},
and in its minimal form it leaves neutrinos
massless~\cite{Valle:2015pba}.

In addressing those questions, it is natural to consider extending the
ideas of gauge symmetry and its spontaneous breaking beyond their
established, central role in the Standard Model.  SU(3) and SO(3) suggest
themselves as candidate symmetries for family unification, since they
support irreducible triplet representations.  (Discrete symmetries can
also be gauged, and quantum gravity might require that they are~\cite{Krauss:1988zc,deMedeirosVarzielas:2017sdv}.)
SO(3) is particularly attractive, since it arises naturally in the
context of comprehensive unification, which brings together forces and
flavor
\cite{GellMann:1980vs,Wilczek:1981iz,Bagger:1984rk,Giudice:1991sz,Reig:2017nrz}.

In this letter we explore an SO(3) family symmetry model inspired by
comprehensive unification.  Within a reasonably economical model,
several appealing features emerge:
\begin{itemize}
\item	A Peccei-Quinn symmetry, leading to axions, which is both natural and helpful to ensure correct mass relations  
 
\item Extreme fine tuning is not requied

\item A characteristic ``golden'' formula relating quark and lepton
  masses, given in Eq.~(\ref{eq:gold}),~\cite{Morisi:2011pt,Morisi:2013eca,Bonilla:2014xla} 
\item A successful explanatory framework for the Cabibbo-Kobayashi-Maskawa (CKM) matrix, with two predictions, Eqs.~(\ref{eq:cab})(\ref{eq:ub}) 
\item A conventional seesaw
  mechanism~\cite{GellMann:1980vs}~\cite{Minkowski:1977sc,Yanagida:1979as,glashow1980future,mohapatra:1980ia,Schechter:1980gr,Lazarides:1980nt} for neutrino mass generation at the Peccei-Quinn (PQ) scale, supplemented by a connection between lepton number and PQ
  breaking, which relates the axion and neutrino mass scales,
  Eq.~(\ref{eq:a-nu})
\end{itemize}

\section{Model construction}
\label{sec:model-construction}

\subsection{SO(3) as family symmetry}
\label{sec:origin-so3_f-family}

Discrete~\cite{Morisi:2011pt,Morisi:2013eca,Bonilla:2014xla} and
continuous~\cite{Wilczek:1978xi,King:2003rf} horizontal flavor
symmetries have been used extensively in model building.  Many options have been considered. It is interesting to consider, as a source of guidance, their possible deeper origin.
        In \cite{Reig:2017nrz} we revived the idea of comprehensive unification, merging gauge and family symmetry.   The striking fact that one can accommodate the observed fermions into a single irreducible spinor multiplet of large orthogonal groups encourages such ideas, which have a long history - see, e.g., ~\cite{Wilczek:1981iz}.   On the face of it, however, they contain too many families, and also an equal number of wrong-chirality ``antifamilies'' (but no other exotics).   We suggested that extraneous families are  confined at a high scale $\gsim$ 10 TeV,  while the antifamilies were removed through an orbifold construction.

        More specifically, the breaking scheme $SO(18) \rightarrow SO(10) \times SO(5) \times SO(3)$ (see
        \cite{GellMann:1980vs}) allows for the standard, attractive, $SO(10)$ gauge unification, together with a hypercolor $SO(5)$ which confines 5 families (leaving 3) and an $SO(3)$ family symmetry group.  This motivates consideration of $SO(3)$ as a family unification group.

More generally, $SO(3)$ family symmetry is more easily compatible with gauge unification than is $SU(3)$ family symmetry.  In the usual
$SU(5)$ and $SO(10)$ theories one embeds the \sm particle content in the
anomaly free sets of representations:
$3\times (\mathbf{\bar{5}}+\mathbf{10})$ for $SU(5)$ and 
$3\times\mathbf{16}$ for $SO(10)$. 
Assigning these representations as $SU(3)$ triplets generally leads to anomalies. For example, in the $SO(10)\times SU(3)$ theory the standard 
 $(\mathbf{16},\mathbf{3})$ combination has an
$[SU(3)_F]^3$ anomaly.  

\subsection{Field content}
\label{sec:model}
We now develop a consistent flavor extension of the Standard Model in which the gauge
symmetry is enlarged by adding the local $SO(3)_F$ family symmetry
\cite{Wilczek:1978xi}.
In addition to \sm particles, the model has an enlarged scalar sector and right handed neutrinos.  This minimal extension is enough to accommodate fermion masses and mixings without fine tuning of parameters, and the other features mentioned earlier.

The field content of our model is displayed in
Table~\ref{tab:content}.  Especially noteworthy are the Peccei-Quinn
charge assignments\footnote{Recently, an alternative framework with flavor-dependent Peccei-Quinn charges has been proposed in~\cite{Ema:2016ops,Calibbi:2016hwq}. Our $U(1)_{PQ}$ symmetry is related to flavor in a rather different way, through the SO(3) family symmetry.}.  They arise from a transformation that commutes
with $SO(10)$: all the fermion fields which occur in the $SO(10)$
spinor have the same PQ charge.  It also commutes with $SO(3)_F$.

\begin{table}[!h]
\begin{center}
\begin{tabular}{|c||c|c|c||c|c|c||c|c||c|c||c|c|}
\hline
 & $q_L$ &$u_R$&$d_R$ &$l_L$ &$e_R$ &$\nu_R$ & $\Phi^u$ & $\Phi^d$ & $\Psi^u$ & $\Psi^d$ & $\sigma$ & $\rho$\\
\hline
$\mathrm{SU(3)_c}$ & \three &\three &\three &\one &\one &\one  & \one & \one & \one &\one & \one & \one \\
\hline
$\mathrm{SU(2)_L}$ & \two & \one & \one  & \two & \one & \one & \two & \two & \two & \two  &\one &\one\\
\hline
$\mathrm{U(1)_{Y}}$ & $\frac{1}{6}$ & $\frac{2}{3}$  & -$\frac{1}{3}$  & -$\frac{1}{2}$ & $-1$ & $0$ & -$\frac{1}{2}$ & $\frac{1}{2}$ & -$\frac{1}{2}$& $\frac{1}{2}$  &$0$ & $0$\\ 
\hline
$\mathrm{SO(3)_F}$ & \three &\three &\three &\three &\three & \three & \five & \five & \three & \three  &$\mathbf{5}$&$\one$\\
\hline
$\mathrm{U(1)_{PQ}}$ & 1 & -1 & -1 & {1} & {-1} & {-1} & 2 & 2 & 2 & 2 & {2} & {2}\\
\hline
\end{tabular}
\caption{Particle content and transformation properties under the \SM
  and flavor SO(3) gauge groups. The VEVs of \SM singlets $\sigma$ and
  $\rho$ break $U(1)_{PQ}$ and lepton number, generating Majorana
  neutrino masses.}
\label{tab:content}
\end{center}
\end{table}

\subsection{Symmetry breaking}
\label{sec:yukawa-lagr-golden}
In our model symmetry breaking proceeds through the following set of
scalar fields: 
\begin{equation}
\begin{split}
&\Psi^u\sim (\one,\two,-1/2,\three)\,,\\&
\Psi^d\sim (\one,\two,1/2,\three)\,,\\&
\Phi^u\sim (\one,\two,-1/2,\five)\,,\\&
\Phi^d\sim (\one,\two,1/2,\five)\,,\\&
\sigma\sim (\one,\one,0,\five)\,,\\&
\rho\sim (\one,\one,0,\one)\,.
\end{split}
\end{equation}
There are two \SM singlet scalars, $\sigma\sim (\one,\one,0,\five)$
and $\rho\sim (\one,\one,0,\one)$.  All of these fields will acquire
nontrivial vacuum expectation values.

Both SO(3) singlet as well as the quintuplet, carry nontrivial PQ
charges. Therefore, the spontaneous breaking of the Peccei-Quinn
symmetry is triggered by their large vacuum expectation values (VEVs).
On the other hand, the SO(3) family symmetry breaking is associated to
the VEV of the \SM singlet scalar ${\sigma}$;
\begin{equation}
\vev{\sigma} = v_\sigma ~\mathrm{diag}(0,1,-1)~~~~~\mathrm{and}~~~~~
\vev{\rho}  = v_\rho ~\mathrm{diag}(1,1,1)
\end{equation}
As we will see later, both VEVs play a key role in breaking lepton
number, generating Majorana neutrino mass, and accounting for the
large neutrino mixing angles observed in neutrino oscillations.

In order to break the electroweak symmetry we assume VEVs for the
$\mathrm{SU(2)_L}$ scalar doublets,
i.e. $\Phi^u$ and $\Phi^d$,  transforming as SO(3) quintuplets, as well as
 $\Psi^u$ and $\Psi^d$,  transforming as  SO(3)  triplets.
We assume the following pattern for the VEVs:
\begin{equation}\label{ssb_3s}
\begin{split}
&\vev{\Phi^{u,d}}=
\left (\begin{array}{ccc}
0&    & \\
& -k^{u,d} & \epsilon_1^{u,d} \\
&  \epsilon_1^{u,d}   &\hspace{1mm}\,\,k^{u,d} \\
\end{array}\right)       \\&
\vev{\Psi^{u,d}}=
\left (\begin{array}{ccc}
v^{u,d} \\
0 \\
\epsilon_2^{u,d} \\
\end{array}\right)~,
\end{split}\end{equation}
where the small parameters parameters $\epsilon_i$ denote a perturbation with respect to the simplest alignments diag(0,-1,1) and (1,0,0).
This symmetry breaking pattern minimizes the Higgs potential~\cite{Wilczek:1978xi}, and provides a good description of the observed fermion mass hierarchy; see below.

An important feature of the model is the existence of a spontaneously broken
U(1) global PQ symmetry. For definiteness, we fix the PQ quantum
numbers as given in Table~\ref{tab:content}. 
The VEVs of \SM singlets $\sigma$ and $\rho$ break $U(1)_{PQ}$ as well
as lepton number. The alignment of the associated Nambu-Goldstone
boson $G$ is
\begin{equation}
\label{eq:profile}
G \approx \frac{1}{(v_\sigma^2 + v_\rho^2)^{1/2}} \left(v_\sigma \sigma^I + v_\rho \rho^I + \dots \right)
\end{equation}
where $\rho^I$ etc. denote the imaginary parts of scalars and $\dots$
denotes components along the isodoublet scalars
$\Psi^{uI}, \Psi^{dI}, \Phi^{uI}, \Phi^{dI}$, weighted by their VEVs
and times their PQ charges.
Notice that, through these projections, $G$ will couple directly to
quarks and leptons at the tree
level. 
These couplings are suppressed linearly by the PQ-breaking scale
$(v_\sigma^2 + v_\rho^2)^{1/2}$.  Thus we arrive at a model of the
DFSZ type~\cite{Dine:1981rt} including coupling to neutrinos.
  
\section{``Golden formula'' for quarks and lepton masses}
\label{sec:gold-form-charg}
Given the SO(3) multiplication rules,
$ \mathbf{3}\times\mathbf{3} =\mathbf{1}+\mathbf{3}+\mathbf{5} \,, $
one can use the vector (triplet) and the two-index symmetric traceless
tensor (quintuplet) representations to build the following invariant
Yukawa Lagrangian,
\begin{equation}\label{eq:yuklag}
\mathcal{L}=\bar{q}_L(y_1\Psi^u+y_2\Phi^u)u_R+\bar{q}_L(y_3\Psi^d+y_4\Phi^d)d_R+\bar{l}_L(y_5\Psi^d+y_6\Phi^d)e_R+h.c.
\end{equation}
Note that the ``duplicated'' scalar sector, with two scalar doublets
selectively coupled to up-type/down-type fermions, does {\it not\/} imply a nonminimal low-energy Higgs sector, as we shall discuss further below.

After electroweak breaking,
Eq.~(\ref{eq:yuklag}) leads to the quark mass matrices
\begin{equation}
\label{mass.u}
M^u=\left(\begin{array}{ccc}
0 & y_1\epsilon^u_2 & 0 \\
-y_1\epsilon^u_2 & -y_2k^u & y_1v^u + y_2\epsilon_1^u \\
0 & -y_1v^u + y_2\epsilon_1^u & y_2k^u
\end{array}\right)
\end{equation}
\begin{equation}
\label{mass.d}
M^d=\left(\begin{array}{ccc}
0 & y_3\epsilon^d_2 & 0 \\
-y_3\epsilon^d_2 & -y_4k^d & y_3v^d + y_4\epsilon_1^d \\
0 & -y_3v^d+ y_4\epsilon_1^d & y_4k^d
\end{array}\right),
\end{equation}
and for the charged leptons 
\begin{equation}
\label{mass.e}
M^e=\left(\begin{array}{ccc}
0 & y_5\epsilon^d_2 & 0 \\
-y_5\epsilon^d_2 & -y_6k^d & y_5v^d+ y_6\epsilon_1^d \\
0 & -y_5v^d+ y_6\epsilon_1^d & y_6k^d
\end{array}\right)~,
\end{equation}
where we take into account the VEV alignment patterns of the SO(3) triplet and quintuplet scalars, respectively.

These matrices allow a good description of the charged fermion masses.
Indeed, neglecting the $\epsilon_i$ parameters, assumed small, 
which describe the departure from the simplest VEV alignment, the
eigenvalues of the matrices are given as~\cite{Wilczek:1978xi}
\begin{equation}\label{masses}
\begin{split}
&m_{u,d, e}=0\,,\\&
m_{c,s,\mu}=|y_{2,4,6}k^{u,d}-y_{1,3,5}v^{u,d}|\,,\\&
m_{t,b,\tau}=|y_{2,4,6}k^{u,d}+y_{1,3,5}v^{u,d}|\,.
\end{split}
\end{equation}
When one takes into account the small perturbations, $\epsilon_i$, one
finds that the ``golden formula'' 
\begin{equation}
	\label{eq:gold}
	\frac{m_\tau}{\sqrt{m_em_\mu}}\approx\frac{m_b}{\sqrt{m_dm_s}}~.
	\end{equation}
        This successful formula nicely relates down-type quark and
        charged lepton masses.  On the other hand, the doubled Higgs structure forced by PQ symmetry allows us to avoid  the unwanted top quark
        mass prediction $\frac{m_\tau}{\sqrt{m_em_\mu}}\approx\frac{m_t}{\sqrt{m_um_c}}$ present in~\cite{Wilczek:1978xi}.
Let us note that the ``golden formula'' relating quark and lepton
masses in Eq.~(\ref{eq:gold}) also emerges in other flavor symmetry
schemes, such as the ones proposed
in~\cite{Morisi:2011pt,Morisi:2013eca,Bonilla:2014xla}, but without connection to an underlying Peccei-Quinn symmetry.

\section{Emergence of the CKM matrix}
\label{sec:gener-ckm-matr}
We now show that, in addition to Eq.~(\ref{eq:gold}), our scheme
provides a dynamical framework for the CKM matrix describing quark
mixing and CP violation.

It is clear from Eqs.~(\ref{mass.u}), (\ref{mass.d}) and (\ref{mass.e})
that, in the limit of vanishing $\epsilon_i$, the charged fermions of
the first family are massless.
Moreover,  when the perturbations $\epsilon_i\to 0$, the
matrix that diagonalizes $M_{u,d}.M^\dagger_{u,d}$ is given by
\begin{equation}\label{eigensystem}
V^{u,d}_L=\left(\begin{array}{ccc}
1 & 0 & 0\\
0 & 1/\sqrt{2} & 1/\sqrt{2} \\
0 & -1/\sqrt{2} & 1/\sqrt{2}\\
\end{array}\right)\,,
\end{equation}
for up and down-type quarks, with eigenvalues given by
Eq. (\ref{masses}). This means that the CKM matrix, defined as
$V_{\text{CKM}}=V_L^uV_L^{d\dagger}$, is naturally ``aligned'' to be
just the identity matrix.

The perturbations of the eigenvectors of $M.M^\dagger$ which result from
turning on the perturbations around the minima get translated into a
small shift of the matrices in Eq.~(\ref{eigensystem}), which no
longer coincide. Their mismatch is the CKM matrix.
After turning on these perturbations, the electron and the up and down
quarks, all get nonzero masses, while small quark mixing angles
emerge naturally.

Thanks to the structured breaking of the $SO(3)$ family symmetry, one
can predict mixing angles in terms of quark masses.  We have the
well-known Gatto-Sartori-Tonin~\cite{Gatto:1968ss} relation for the
Cabibbo angle
\begin{equation}
  \label{eq:cab}
  \theta_C \approx \sqrt{\frac{m_d}{m_s}}-\sqrt{\frac{m_u}{m_c}}~,
\end{equation}
while for $|V_{ub}|$ we get
\begin{equation}
  \label{eq:ub}
  |V_{ub}| \approx \frac{ \sqrt{m_d m_s}}{m_b}- \frac{ \sqrt{m_u m_c}}{m_t}~,
\end{equation}
which extends a relation found in Ref.~\cite{Wilczek:1978xi}.
Finally, the doubling of scalar quintuplets
$\vev{\Phi^{u,d}}$ plays a crucial role in generating $|V_{cb}|$,
given as
\begin{equation}
  \label{eq:cb}
|V_{cb}| = \frac{\epsilon_1^u}{2k^u}-\frac{\epsilon_1^d}{2k^d}\,. 
\end{equation}
In contrast to $\theta_C$ and $|V_{ub}|$, the $|V_{cb}|$ matrix
element can only emerge from the duplicated set of quintuplets,
i.e. from the fact that $\Phi^u$ and $\Phi^d$ are different
fields. Otherwise, the $b$ quark would decay predominantly to up
quarks through the weak charged current.
Thus, in the present framework mass hierarchies and mixing
angles arise as perturbations around the symmetry breaking minima of
the scalar potential, rather than hierarchies in the Yukawa couplings. 
CP violation can be accommodated through nontrivial phases in the
Yukawa couplings, but no useful prediction emerges.

In short, the predictions we found confirm the expectation from a simple parameter counting. Including Yukawa couplings and VEVs we have in total 10 relevant parameters to describe 13 observables, namely the 9 charged fermion masses plus the 4 CKM parameters~\footnote{ 
The neutrino sector is discussed separately, see below.}. This leads to three successful relations. Two of these are the predictions for the charged fermion masses, Eq.~(\ref{eq:gold}), and for the Cabibbo angle, Eq.~(\ref{eq:cab}). The third relation is given in Eq.~(\ref{eq:ub}).
Notice that, thanks to the PQ symmetry, the ``golden'' relation in Eq.~(\ref{eq:gold}) is a successful one, as it involves only the ``down-type'' fermions, in contrast to Ref.~\cite{Wilczek:1978xi}. Likewise, we have that $|V_{cb}|,~|V_{ub}| \neq 0$, as required.
Note also the important role played by the scalar potential dynamics, namely, the need for VEV misalignment in Eq.~(\ref{ssb_3s}). 

\section{Neutrino masses and mixings}
\label{sec:gener-ckm-matr-1}

Neutrino masses arise naturally in $SO(10)$ unification through a conventional (type I) seesaw mechanism~\cite{GellMann:1980vs}~\cite{Minkowski:1977sc,Yanagida:1979as,glashow1980future,mohapatra:1980ia,Schechter:1980gr,Lazarides:1980nt}.
In order to capture this at our level of analysis we add right-handed neutrinos $\nu_R$ transforming under the Peccei-Quinn symmetry, as in Table~\ref{tab:content}.

The relevant Yukawa Lagrangian to generate neutrino masses is given by

\begin{equation}
\label{ssw}
\mathcal{L}_\nu=\bar{l}_L(y_D\Psi^u+\tilde{y}_D\Phi^u)\nu_R+ \bar{\nu}_R^c(y_M\sigma+y^\prime_M\rho)\nu_R\,,
\end{equation}
The vacuum expectation values of $\rho$ and $\sigma$ break the
Peccei-Quinn symmetry spontaneously, as well as lepton
number.
The last terms in Eq.~(\ref{ssw}) generate Majorana masses for $\nu_R$
after symmetry breaking. Notice also that $\vev{\sigma}$ breaks the
SO(3) family symmetry.

To support a viable seesaw mechanism, both $\rho$ and $\sigma$ are
necessary.  If there were only the flavor singlet, neutrino mixing
would be similar to that of quarks, hence small, and ruled out by the
neutrino oscillation data~\cite{deSalas:2017kay}.  Were there only the
quintuplet, a two index traceless symmetric tensor, the seesaw would
be singular, leaving four light neutrinos, instead of three.
The symmetry breaking pattern obtained through the simultaneous
presence of $\sigma$ and $\rho$ plays a key role in order to account
for why neutrinos mix in such a different way from quarks.

In short, in our model the quark mixing and CP violation arise from
departures from the simplest VEV alignment of the Higgs fields,
$\epsilon_i \neq 0$ in Eq.~(\ref{ssb_3s}), and are significantly
constrained.
In contrast, neutrino masses and (generically large) lepton mixing are
directly associated with Peccei-Quinn breaking.


\section{Higgs scalar spectrum}

Our explicit implementation of SO(3) flavor symmetry requires several
scalar multiplets.  In the context of renormalizable quantum field
theory, without further constraints, there are many scalar coupling
terms, and - given that most of the spectrum is lifted to a high mass
scale - few observational handles on them.  Thus a complete analysis
is both impractical and pointless; but we do need to ensure that an
acceptable low-energy sector can emerge.

Generically, all the fields other than the axion will acquire mass
terms of order the flavor and PQ breaking scale, barring cancellations
between bare and induced mass terms.  For purposes of
$SU(2)\times U(1)$ breaking, we require at least one much lighter
doublet.  Notoriously, that requires a conspiracy or fine-tuning among
parameters.  This is an aspect of the hierarchy problem, which we do
not address here.  The only slight good news is that the existence of
more than one doublet would require additional fine tuning, so that
the minimal one doublet structure, which so far is supported by
experimental observations, is minimally unnatural.

To illustrate the mechanism whereby induced mass terms arise, consider the
quartic operator $\Psi_u^\dagger\Psi_u\sigma^\dagger\rho$. Its
contraction is unique and can be easily visualized in matrix form.
If $\vev{\sigma}$ is aligned in the diagonal (recall it is symmetric
and traceless) and $\rho$ is an SO(3) singlet we get, after SO(3)
breaking takes place,
\begin{equation}
\begin{split}
&\vev{\rho} \Psi_u^\dagger\vev{\sigma^\dagger}\Psi_u=
v_\rho\left(\begin{array}{ccc}
\Psi_{1u}^{\dagger}&\Psi_{2u}^{\dagger}&\Psi_{3u}^{\dagger}
\end{array}\right)
\left(\begin{array}{ccc}
0&0&0\\
0&-v_\sigma&0\\
0&0&v_\sigma
\end{array}\right)
\left(\begin{array}{c}
\Psi_{1u}\\\Psi_{2u}\\\Psi_{3u}
\end{array}\right)\\&
=-v_\sigma v_\rho|\Psi_{2u}|^2+v_\sigma v_\rho|\Psi_{3u}|^2\,.
\end{split}
\end{equation}

One sees that the vacuum expectation value of the above operator
  generates a splitting of order the flavor/PQ breaking scale among
  the electroweak doublet components of $\Psi_u$, so that two of them
  can be made heavy, i.e. at the large symmetry breaking scale,
  leaving the other massless.
  This argument may be escalated to the full scalar potential, which
  contains many relevant quartics,  viz.
\begin{eqnarray}
&\Phi_{u}^\dagger\Phi_{u}\sigma^\dagger\sigma\,,\,\,\Phi_{u}^\dagger\Phi_{u}\sigma^\dagger\rho\,,\,\,\Phi_{u}^\dagger\Phi_{u}\rho^\dagger\rho\,, 
\,\,\Psi_{u}^\dagger\Psi_{u}\sigma^\dagger\sigma\,,\,\,\Psi_{u}^\dagger\Psi_{u}\sigma^\dagger\rho\,,\,\,\Psi_{u}^\dagger\Psi_{u}\rho^\dagger\rho\,, \nonumber \\
&\Phi_{d}^\dagger\Phi_{d}\sigma^\dagger\sigma\,,\,\,\Phi_{d}^\dagger\Phi_{d}\sigma^\dagger\rho\,,\,\,\Phi_{d}^\dagger\Phi_{d}\rho^\dagger\rho\,,\,\,\Psi_{d}^\dagger\Psi_{d}\sigma^\dagger\sigma\,,\,\,\Psi_{d}^\dagger\Psi_{d}\sigma^\dagger\rho\,,\,\,\Psi_{d}^\dagger\Psi_{d}\rho^\dagger\rho\,,
\nonumber \\
&\Phi_{u}^\dagger\Psi_{u}\sigma^\dagger\sigma\,,\,\,\Phi_{u}^\dagger\Psi_{u}\sigma^\dagger\rho\,,\,\,\Phi_{u}^\dagger\Psi_{u}\sigma\rho^\dagger\,,
\Phi_{d}\Psi_{d}^\dagger\sigma^\dagger\sigma\,,\,\,\Phi_{d}\Psi_{d}^\dagger\sigma^\dagger\rho\,,\,\,\Phi_{d}\Psi_{d}^\dagger\sigma\rho^\dagger\,, 
\nonumber \\ 
&\Phi_{u}\Phi_{d}\sigma^\dagger\sigma^\dagger\,,\,\,\Phi_{u}\Phi_{d}\sigma^\dagger\rho^\dagger\,,\,\,\Phi_{u}\Phi_{d}\rho^\dagger\rho^\dagger\,,\,\,\Psi_{u}^\dagger\Psi_{d}^\dagger\sigma\sigma\,,\,\,\Psi_{u}^\dagger\Psi_{d}^\dagger\sigma\rho\,,\,\,\Psi_{u}^\dagger\Psi_{d}^\dagger\rho\rho\,, \nonumber \\ 
&\Phi_{d}\Psi_{u}\sigma^\dagger\sigma^\dagger\,,\,\,\Phi_{d}\Psi_{u}\sigma^\dagger\rho^\dagger\,,\,\,\Phi_{u}^\dagger\Psi_{d}^\dagger\sigma\sigma\,,\,\,\Phi_{u}^\dagger\Psi_{d}^\dagger\sigma\rho\,
\end{eqnarray}
(These can be obtained in a
systematic way; see ~\cite{Fonseca:2017lem}.)
One finds that, after breaking, the scalar mass$^2$ matrix typically
contains suitable off-diagonal terms, ensuring that the light doublet
is a linear combination of $\Psi^u, \Psi^d, \Phi^u, \Phi^d$ wherein
each appears with a nonzero coefficient.

\section{Discussion} 

Before closing, we comment briefly on three issues which deserve mention.
\begin{enumerate}

\item The PQ symmetry $U(1)_{PQ}$ is conserved at the classical level,
  and to all orders in perturbation theory, but violated
  nonperturbatively.  One can visualize the breaking using QCD
  instantons, and infer its character by analyzing anomalies.  In this
  way, one may discover that a nontrivial $Z_N$ subgroup of
  $U(1)_{PQ}$ is valid even nonperturbatively.  Our model, as it
  stands, has $N = 12$, with doubly charged scalar fields.  If scalar
  fields which are not $Z_N$ singlets acquire VEVs, the possibility of
  domain walls arises.  Such domain walls are very dangerous for early
  universe cosmology~\cite{Zeldovich:1974uw}.
  The most straightforward way to avoid this difficulty is to assume
  that the $Z_N$ breaking is followed by a period of cosmic inflation,
  so that potential domain walls get pushed beyond the horizon.
  Another possibility is to arrange that $N = 1$.  (We could also
  allow $N =2$, since the PQ-breaking VEVs have PQ charge 2). This
  does not occur in our model as it stands, but it can be achieved by
  adding suitable colored fermions.  In the absence of other
  motivations, however, that construction seems contrived.

\item The vacuum expectation values of the $\rho$ and $\sigma$ scalars
  are responsible both for Peccei-Quinn and lepton number symmetry
  breaking.
This entails an interesting conceptual relation between the axion and
neutrino mass scales, of the form
\begin{equation}
\label{eq:a-nu}
m_a \sim (\Lambda_{QCD}m_\pi / v^2)m_\nu \,.
\end{equation}
where $m_\pi$ is the pion mass and $v$ is the electroweak scale.  This
relation, which implies that the axion mass is parametrically smaller
than the neutrino mass, according to the square of the ratio of QCD to
electroweak scales.  Since it assumes that the Yukawa couplings
involving the neutrino field is of order unity, it should be applied
using the heaviest of the light neutrinos.  Of course, we
cannot preclude the possibility that PQ symmetry breaks at a higher
scale, through condensates which do not generate neutrino masses; this
effect could drive the axion mass down further.


\item The presence of extra gauge bosons coupled to flavor will
  mediate $\Delta F=2$ neutral flavor changing interactions at
  tree-level.  The most sensitive probe appears to be $K^0-\bar{K}^0$
  mixing~\cite{Patrignani:2016xqp}.  From this we estimate
\begin{equation}
\frac{g^2}{M_F^2}\lsim \frac{1}{[10^4\, \text{TeV}]^2}\,,
\end{equation}
where the gauge boson mass is $M_F\sim g f_a$. This constrains the
Peccei-Quinn breaking scale to be $f_a\gsim 10^7$ GeV, a much weaker
bound than arises from astrophysical
constraints~\cite{Bertolami:2014wua}.

\end{enumerate}

\section{Summary}

Motivated by ideas arising in comprehensive unification based on
spinors, we have considered possible consequences of supplementing the
\sm gauge symmetry with commuting $SO(3)$ flavor and PQ symmetries in
a way consistent with $SO(10)$ embedding.
Proceeding in a bottom-up way, we analyzed a minimal
$SO(3)_F \times U(1)_{PQ}$ extension of the \sm unifying together the
three families of matter.  Fairly simple choices of multiplet
structure and symmetry breaking pattern allowed us to accommodate the
known phenomenology of quark and lepton masses and mixings and to make
several nontrivial connections among them.  The PQ symmetry was
important to this success, and of course it continues to serve its
familiar roles in ensuring accurate strong $T$ symmetry and in providing,
in axions, a good dark matter candidate.

\section{Acknowledgements}

This work is funded by the Spanish grants SEV-2014-0398 and FPA2017-85216-P (AEI/FEDER, UE) and PROMETEO/2018/165 (Generalitat Valenciana). The work of M.R. is supported by FPU grant FPU16/01907.  F.W.'s work is supported by the U.S. Department of Energy under grant Contract  Number DE-SC0012567, by the European Research Council under grant No. 742104, and by the Swedish Research Council under Contract No. 335-2014-7424. M.R. would like to thank Aqeel Ahmed for helpful conversations.


\providecommand{\href}[2]{#2}\begingroup\raggedright\endgroup

\end{document}